\documentclass[twocolumn,tighten,pdfLaTeX]{aastex6}
\usepackage{natbib}

\begin{document}


\title{Ejection of Chondrules from Fluffy Matrices
}

%% Use \author, \affil, plus the \and command to format author and affiliation 
%% information.  If done correctly the peer review system will be able to
%% automatically put the author and affiliation information from the manuscript
%% and save the corresponding author the trouble of entering it by hand.
%%
%% The \affil should be used to document primary affiliations and the
%% \altaffil should be used for secondary affiliations, titles, or email.

%% Authors with the same affiliation can be grouped in a single
\author{Sota Arakawa}
\affil{Department of Earth and Planetary Sciences, Tokyo Institute of Technology, Meguro, Tokyo, 152-8551, Japan}

%% Use the \and command so offset the last author.


\begin{abstract}
Chondritic meteorites primarily contain millimeter-sized spherical objects, the so-called chondrules; however, the co-accretion process of chondrules and matrix grains is not yet understood.
In this study, we investigate the ejection process of chondrules via collisions of fluffy aggregates composed of chondrules and matrices.
We reveal that fluffy aggregates cannot grow into planetesimals without losing chondrules if we assume that the chondrite parent bodies are formed via direct aggregation of similar-sized aggregates.
Therefore, an examination of other growth pathways is necessary to explain the formation of rocky planetesimals in our solar system.
\end{abstract}




\section{Introduction}

Terrestrial planets are believed to form in protoplanetary disks via collisions and the coalescence of rocky planetesimals; however, the process by which interstellar dust grains evolve into planetesimals is still enigmatic.
This is because there are several ``barriers'' to planetesimal formation, e.g., the bouncing barrier \citep[e.g.,][]{Zsom+2010}, the fragmentation barrier \citep[e.g.,][]{Blum+2008}, and the radial drift barrier \citep[e.g.,][]{Weidenschilling1977}.

The classical proposed mechanism for planetesimal formation is the gravitational instability of thin dust layer \citep[e.g.,][]{Goldreich+1973,Sekiya1983}; however, it is now believed that the Kelvin-Helmholtz instability generated by vertical shears and the differential motions of dust particles and gas prevents disks from entering a gravitationally unstable state \citep[e.g.,][]{Cuzzi+1993}.
Subsequently, several alternative scenarios have been proposed to explain planetesimal formation, for example, the streaming instability \citep[e.g.,][]{Youdin+2005,Johansen+2007} and collisional growth via direct aggregation \citep[e.g.,][]{Windmark+2012,Kataoka+2013b}.

The streaming instability is driven by differences in the motions of the gas and dust particles in the disk, and this mechanism requires a large dust-to-gas ratio \citep[e.g.,][]{Drazkowska+2014}.
In addition, the streaming instability requires grown dust aggregates whose motions are marginally decoupled from the disk gas \citep[e.g.,][]{Bai+2010}.
Therefore, whether planetesimals can form via the streaming instability depends on the vertically integrated dust-to-gas mass ratio and the critical velocity for collisional growth.

The direct aggregation scenario is also restricted by the solid mass fraction and the critical velocity for growth.
At least for aggregates composed of submicron ice grains, the critical velocity could exceed the maximum collisional velocity of dust aggregates in protoplanetary disks \citep[e.g.,][]{Wada+2009,Wada+2013}.
In addition, if the fluffy growth of dust aggregates is taken into consideration, \citet{Okuzumi+2012} have shown that these aggregates can avoid the radial drift barrier even if they grow in the minimum mass solar nebula \citep{Hayashi1981}.
Moreover, according to numerical simulations, fluffy aggregates are less likely to bounce when they collide \citep{Wada+2011,Seizinger+2013}.

By contrast, the critical collision velocity of rocky aggregates is still open to argument.
While the critical velocity strongly depends on the surface energy and the radius of the monomers \citep{Dominik+1997}, both the surface energy \citep[e.g.,][]{Yamamoto+2014,Kimura+2015} and the monomer radius \citep{Arakawa+2016b} are still under debate.
Additional mechanisms for increasing the critical velocity by dissipating the kinetic energy when dust aggregates collide have also been suggested by \citet{Tanaka+2012} and \citet{Krijt+2013}.
Furthermore, the critical velocity increases if the silicate monomers are coated with sticky organic mantles \citep[e.g.,][]{Piani+2017} or if the collisions of the rocky aggregates occur in high-temperature environments \citep[e.g.,][]{Hubbard2017}.

If we discuss the formation process of rocky planetesimals in our solar system, we must consider how the chondrite parent bodies formed.
Chondrites are one of the major groups of meteorites, and it has been shown that S-type asteroids are the parent bodies of ordinary chondrites \citep{Nakamura+2011}.
Ordinary chondrites primarily consist of millimeter-sized spherical objects so-called chondrules and nanometer- and micron-sized matrix grains between the chondrules \citep[][and references therein]{Scott2007}.
Therefore, rocky planetesimals in our solar system must be formed via the co-accretion of chondrules and matrix grains; however, it is not yet understood how chondrite parent bodies are made.

There are a few studies that have investigated the co-accretion process of chondrules and matrices.
\citet{Ormel+2008} have shown that free-floating chondrules in a protoplanetary disk can obtain a porous dust layer around them, which absorbs the kinetic energy of collisions.
\citet{Beitz+2012} also suggested that chondrite parent bodies could have formed from clusters of dust-rimmed chondrules based on low-velocity collisional experiments using millimeter-sized glass beads coated with micron-sized silica particles.
In addition, granular-mechanics simulations \citep{Gunkelmann+2017} revealed that dust layers around chondrules lead to a significant increase of the sticking velocity.

However, if fluffy aggregates collide at large velocities, the chondrules contained in the fluffy matrices might move in the aggregates.
Moreover, a fluffy aggregate cannot retain its chondrules if the length for stopping the chondrules exceeds the radius of the aggregate.
This ejecting of chondrules from the fluffy matrices has the potential to prevent chondritic planetesimal formation.
Therefore, in this study, we examine whether chondrules can be retained inside fluffy aggregates when aggregates collide and grow in a protoplanetary disk.

\section{Models}

\subsection{Outline}

\begin{figure}[h]
\includegraphics[width=\columnwidth]{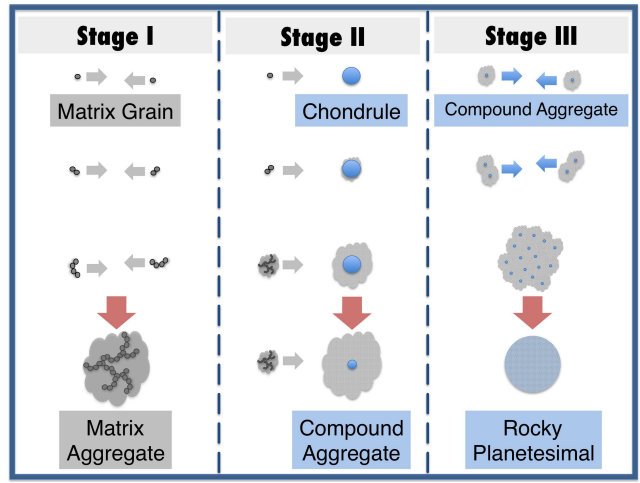}
\caption{Outline of the chondritic rocky planetesimal formation scenario.
Initially monomers of the matrix grains grow and turn into fluffy MAs (Stage I), then CAs form via the accretion of MAs onto chondrules (Stage \mbox{I\hspace{-.1em}I}).
Therefore, matrix grains experience hierarchical growth.
Subsequently, CAs formed in Stage \mbox{I\hspace{-.1em}I} must grow into chondritic rocky planetesimals without losing chondrules (Stage \mbox{I\hspace{-.1em}I\hspace{-.1em}I}).
In this study, we examine the direct aggregation scenario assuming that the density of the CAs is determined by static compression processes.
}
\label{fig1}
\end{figure}

The goal of this paper is to explore whether fluffy aggregates composed of chondrules and matrix grains can grow into chondrite parent bodies without chondrule losses.
Therefore, it is necessary to calculate the structure of the dust aggregates, such as the radius and the bulk density, and the collision velocity.
Here we assume that there are two types of aggregates, that is, chondrule-dust compound aggregates (hereinafter referred to as {\it compound aggregates} or {\it CAs}) and aggregates that are composed purely of matrix grains (hereinafter referred to as {\it matrix aggregates} or {\it MAs}).
The formation processes of MAs and CAs are illustrated in Figure \ref{fig1}.

In Stage I, the growth of MAs primarily occurs owing to Brownian motion, and they reduce their internal densities via fractal aggregation.
We define Stage I as where the timescale of collisions between two MAs $t_{{\rm MA} \to {\rm MA}}$ is shorter than the timescale for MAs to accrete onto chondrules/CAs $t_{{\rm MA} \to {\rm CA}}$.
Then, in Stage \mbox{I\hspace{-.1em}I}, the masses of the CAs increase via the accretion of the grown MAs, and finally, in Stage \mbox{I\hspace{-.1em}I\hspace{-.1em}I}, the growth of CAs via the collision of two CAs with static compression takes place.
We also define Stage \mbox{I\hspace{-.1em}I\hspace{-.1em}I} as where the growth timescale of CAs by the collision of two CAs $t^{{\rm CA} \gets {\rm CA}}_{\rm grow}$ is shorter than the growth timescale of CAs by the accretion of the grown MAs $t^{{\rm CA} \gets {\rm MA}}_{\rm grow}$.
The growth timescale of CAs by the collision of two CAs $t^{{\rm CA} \gets {\rm CA}}_{\rm grow}$ is the same as the timescale of collisions between two CAs $t_{{\rm CA} \to {\rm CA}}$.
On the other hand, the growth timescale of CAs by the accretion of the grown MAs is $t^{{\rm CA} \gets {\rm MA}}_{\rm grow} = {(m_{\rm CA} / m_{\rm MA})} t_{{\rm CA} \to {\rm MA}}$, where $m_{\rm MA}$ and $m_{\rm CA}$ are the masses of MAs and CAs, respectively, and $t_{{\rm CA} \to {\rm MA}}$ is the timescale for CAs to collide with MAs.
We can imagine that fine-grained matrices could help CAs grow in protoplanetary disks due to their large adhesive force.
Forming highly porous matrices also encourages CAs to grow without significant radial drift because fluffy dust aggregates have large collisional cross sections and can grow rapidly.

In this study, we focus on Stage \mbox{I\hspace{-.1em}I\hspace{-.1em}I} and only consider the existence and growth of CAs for simplicity.
We calculate the density evolution of the CAs by considering the compression processes and examine whether the CAs can retain chondrules inside their fluffy matrices.

\subsection{Structure of the Disk}

We assume that the disk structure is in steady state, i.e., the mass accretion rate $\dot{M}$ is constant with distance from the Sun $R$.
Because a large amount of fine dust particles must have been present when planetesimals formed in the solar nebula, we also assume that the disk is optically thick and that the main heating source at the midplane is viscous dissipation.
Then the local surface density $\Sigma$ and the midplane temperature $T_{\rm m}$ are given by \citep[e.g.,][]{Pringle1981,Oka+2011}
\begin{equation}
\Sigma = \frac{\dot{M}}{3 \pi \nu_{\rm acc}},
\end{equation}
and
\begin{equation}
T_{\rm m} = {\left( \frac{9 \kappa \Sigma}{32} \frac{G M_{\odot} \dot{M}}{4 \pi \sigma_{\rm SB} R^{3}} \right)}^{1/4},
\end{equation}
where $\nu_{\rm acc}$ is the effective viscosity of the accretion disk, $\kappa$ is the Rosseland mean opacity of the disk medium, $G = 6.67 \times 10^{-8}\ {\rm dyn}\ {\rm cm}^{2}\ {\rm g}^{-2}$ is the gravitational constant, and $\sigma_{\rm SB} = 5.67 \times 10^{-5}\ {\rm erg}\ {\rm cm}^{-2}\ {\rm K}^{-4}\ {\rm s}^{-1}$ is the Stefan-Boltzmann constant.
The solar mass $M_{\odot}$ is $M_{\odot} = 1.99 \times 10^{33}\ {\rm g}$.
The Kepler frequency $\Omega_{\rm K}$ is given by $\Omega_{\rm K} = \sqrt{{G M_{\odot}} / R^{3}}$ and the isothermal sound velocity $c_{\rm s}$ is given by $c_{\rm s} = \sqrt{k_{\rm B} T_{\rm m} / m_{\rm g}}$, where $k_{\rm B} = 1.38 \times 10^{-16}\ {\rm erg}\ {\rm K}^{-1}$ is the Boltzmann constant and $m_{\rm{g}}$ is the mean molecular weight.
We set the mean molecular weight to $m_{\rm g} = 2.34 m_{\rm H}$, where $m_{\rm H} = 1.67 \times 10^{-24}\ {\rm g}$ is the mass of a hydrogen atom.
Then, the effective viscosity $\nu_{\rm acc}$ is written as $\nu_{\rm acc} = \alpha_{\rm acc} {c_{\rm s}}^{2} / \Omega_{\rm K}$ \citep{Shakura+1973}.
Here, we assume $\alpha_{\rm acc} = 10^{-2}$ in this study \citep[e.g.,][]{Hartmann+1998}.

The surface densities of dust and gas, $\Sigma_{\rm d}$ and $\Sigma_{\rm g}$, respectively, are given by $\Sigma_{\rm d} = Z \Sigma$ and $\Sigma_{\rm g} = {( 1 - Z )} \Sigma$, where $Z$ is the vertically integrated dust-to-total mass ratio.
The vertical structure of the gas is assumed to be in hydrostatic equilibrium, and the midplane gas density $\rho_{\rm g}$ is given by $\rho_{\rm g} = \Sigma_{\rm g} / {(\sqrt{2 \pi} h_{\rm g})}$, where $h_{\rm g} = c_{\rm s} / \Omega_{\rm K}$ is the gas scale height.
The mean free path of gas molecules $\lambda_{\rm m}$ is given by $\lambda_{\rm m} = m_{\rm g} / {(\sigma_{\rm mol} \rho_{\rm g})}$, where $\sigma_{\rm mol} = 2 \times 10^{-15}\ {\rm cm}^{2}$ is the collisional cross section of the gas molecules \citep{Okuzumi+2012}.
The opacity of the disk $\kappa$ is determined by the surface density of the fine dust, i.e., the matrix grains and small MAs.
We assume that the mass fraction of matrix grains to all solids $\chi$ is equal in the disk and in the CAs.
According to \citet{Scott2007}, the volume fraction of the matrices in ordinary chondrites is approximately 10\%, and the mass fraction is also approximately 10\%.
Therefore, we set $\chi = 0.1$ as the canonical value and the opacity of the disk is given by $\kappa = \chi Z \kappa_{\rm d}$, where $\kappa_{\rm d} = 10^{3}\ {\rm cm}^{2}\ {\rm g}^{-1}$ is the mass opacity of the fine silicate grains \citep[e.g.,][]{Draine1985}.
The mass opacity of highly porous aggregates composed of fine silicate grains is also on the order of $10^{3}\ {\rm cm}^{2}\ {\rm g}^{-1}$ \citep{Kataoka+2014}.

\subsection{Dust Dynamics in the Disk}

We consider the Brownian motion, turbulence, and radial and azimuthal drift as sources of the relative velocity of the gas and dust particles.
We write the relative velocity $v$ as the root sum square of these contributions:
\begin{equation}
v = \sqrt{{v_{\rm{B}}}^{2} + {v_{\rm{t}}}^{2} + {v_{\rm{r}}}^{2} + {v_{\phi}}^{2}},
\end{equation}
where $v_{\rm{B}}$, $v_{\rm{t}}$, $v_{\rm{r}}$, and $v_{\phi}$ are the relative velocities induced by Brownian motion, turbulence, radial drift, and azimuthal drift, respectively.
In this study, we assume that the dust aggregates have no mass distribution and that aggregates that have the same mass have the same volume for simplicity.
The collision velocity $\Delta v$ is
\begin{equation}
{\Delta v} = \sqrt{{(\Delta v_{\rm{B}})}^{2}+ {(\Delta v_{\rm{t}})}^{2}},
\end{equation}
where $\Delta v_{\rm{B}}$ and $\Delta v_{\rm{t}}$ are the collisional velocities induced by the Brownian motion and turbulence, respectively.
The model of the dust dynamics is the same as that of \citet{Arakawa+2016b}.

\subsubsection{Brownian Motions}

The Brownian-motion-induced velocities $v_{\rm{B}}$ and $\Delta v_{\rm{B}}$ are given by 
\begin{equation}
v_{\rm{B}} = \sqrt{\frac{8 k_{\rm{B}} T}{\pi m}},
\end{equation}
and
\begin{equation}
\Delta v_{\rm{B}} = \sqrt{\frac{16 k_{\rm{B}} T}{\pi m}},
\end{equation}
where $m$ is the mass of an aggregate.

\subsubsection{Turbulent Motions}

The dynamics of dust particles in a disk is affected by disk gas turbulence.
The turnover time and the root-mean-squared random velocity of the largest turbulent eddies, $t_{\rm L}$ and $v_{\rm L}$, respectively, are given by $t_{\rm L} = {\Omega_{\rm K}}^{-1}$ and $v_{\rm L} = \sqrt{\alpha_{\rm t}} c_{\rm s}$, where $\alpha_{\rm t}$ is the dimensionless turbulent parameter at the midplane \citep{Cuzzi+2003}.
The value of the dimensionless parameter $\alpha_{\rm t}$ for our solar nebula is unclear; however, measurements of CO emission lines reveal that some circumstellar disks, such as the disk around HD 163296, have a small turbulent parameter \citep[$\alpha_{\rm t} \lesssim 10^{-3}$][]{Flaherty+2015}.
The disk around HL Tau also has a small turbulent viscosity, which is equivalent to an $\alpha_{\rm t}$ at of a few $10^{-4}$ when considering the dust settling suggested by observations of the gap structure \citep{Pinte+2016}; meanwhile, $\alpha_{\rm acc} \sim 10^{-2}$ is indicated by the disk mass distribution and the accretion rate \citep[$8.7 \times 10^{-8}\ M_{\odot}\ {\rm yr}^{-1}$;][]{Beck+2010}.
If the magnetorotational instability is active, $\alpha_{\rm t}$ may be up to $10^{-3}$ \citep[e.g.,][]{Balbus+1991}, whereas $\alpha_{\rm t}$ might be lower than $10^{-3}$ to $10^{-5}$ if we focus on the low-ionization-fraction regions \citep[e.g.,][]{Gammie1996,Mori+2016}.
In this study, we assume that the turbulence is weak in the midplane of the inner disk and that $\alpha_{\rm t} = 10^{-4}$.
The turbulent viscosity $\nu_{\rm t}$ is given by $\nu_{\rm t} = \alpha_{\rm t} c_{\rm s} h_{\rm g}$.
The extent of the gas turbulence is determined by the turbulent Reynolds number ${\rm Re}_{\rm t}$.
The turbulent Reynolds number ${\rm Re}_{\rm t}$ is given by ${\rm Re}_{\rm t} = \nu_{\rm t} / \nu_{\rm m}$, where $\nu_{\rm m} = \lambda_{\rm m} v_{\rm th} / 2$ is the molecular viscosity, and $v_{\rm th} = \sqrt{{8}/{\pi}} c_{\rm s}$ is the thermal velocity of the gas molecules.
The turnover time and the root-mean-squared random velocity of the smallest eddies, $t_{\eta}$ and $v_{\eta}$, respectively, are given by $t_{\eta} = {\rm Re}_{\rm t}^{- 1/2} t_{\rm L}$ and $v_{\eta} = {{\rm Re}_{\rm t}}^{- 1/4} v_{\rm L}$.

The key parameter that determines the motion of dust aggregates in the gas disk is the normalized stopping time, the so-called the Stokes number ${\rm St}$, given by
\begin{equation}
{\rm St} = \Omega_{\rm K} t_{\rm s},
\end{equation}
where $t_{\rm s}$ is the stopping time of the dust aggregates.
The stopping time depends on the radius of the aggregates $a$ and the particle Reynolds number ${\rm Re}_{\rm p} = {2 a v} / \nu_{\rm m}$.
The stopping time $t_{\rm s}$ is given by \citet{Probstein+1970} and \citet{Weidenschilling1977} as follows.
If the radius of an aggregate $a$ is smaller than ${(9 / 4)} \lambda_{\rm m}$, then the stopping time of the dust aggregates is determined by Epstein's law, $t_{\rm s}^{\rm Ep} = {3 m} / {(4 \pi \rho_{\rm g} v_{\rm th} a^{2})}$.
Conversely, when the radius $a$ is larger than ${(9 / 4)} \lambda_{\rm m}$, the stopping time depends on the particle Reynolds number.
If the particle Reynolds number ${\rm Re}_{\rm p}$ is less than unity, the motion of the aggregate is determined by Stokes's law, $t_{\rm s}^{\rm St} = {m} / {(6 \pi \rho_{\rm g} \nu_{\rm m} a)}$, while for the case of $1 \leq {\rm Re}_{\rm p} < 54^{5/3} \sim 800$, the stopping time is obtained from Allen's law, $t_{\rm s}^{\rm Al} = {(2^{3/5} m)} / {(12 \pi \rho_{\rm g} {\nu_{\rm m}}^{3/5} v^{2/5} a^{7/5})}$.
Finally, for the case where the aggregate moves fast and the particle Reynolds number is larger than $54^{5/3}$, we use Newton's law to obtain the stopping time, $t_{\rm s}^{\rm Nw} = {(9 m)} / {(2 \pi \rho_{\rm g} v a^{2})}$.

We use the analytic formulae derived by \citet{Ormel+2007} for the turbulence-driven velocity.
If the stopping time $t_{\rm s}$ is shorter than the turnover time of the smallest eddies $t_{\eta}$, the relative velocity between the dust particle to disk gas $v_{\rm t}$ is given by $v_{\rm t}^{\rm S}$ such that
\begin{equation}
v_{\rm t}^{\rm S} = {{\rm Re}_{\rm t}}^{1/4} {\rm St} v_{\rm L}.
\end{equation}
For the case of $t_{\eta} \ll t_{\rm s} \ll t_{\rm L}$, the relative velocity $v_{\rm t}$ is given by $v_{\rm t}^{\rm M}$ such that
\begin{equation}
v_{\rm t}^{\rm M} =  1.7 {\rm St}^{1/2} v_{\rm L},
\label{eq9}
\end{equation}
and for the case of $t_{\rm s} \gg t_{\rm L}$, the relative velocity $v_{\rm t}$ is given by $v_{\rm t}^{\rm L}$ such that
\begin{equation}
v_{\rm t}^{\rm L} = \left( 1 + \frac{1}{1 + {\rm St}} \right)^{1/2} v_{\rm L}.
\end{equation}
Here, we assume that the turbulence-driven velocity $v_{\rm t}$ is given by the minimum of these three terms for continuity and better handling of $v_{\rm t}$:
\begin{equation}
v_{\rm t} = \min{\left( v_{\rm t}^{\rm S}, v_{\rm t}^{\rm M}, v_{\rm t}^{\rm L} \right)}.
\end{equation}

The turbulence-driven velocity between two dust particles was also obtained by \citet{Ormel+2007}.
We assume the collision velocity of two particles with a stopping time $t < t_{\eta}$ to be
\begin{equation}
\Delta v_{\rm t}^{\rm S} = \varepsilon {{\rm Re}_{\rm t}}^{1/4} {\rm St} v_{\rm L},
\label{eq12}
\end{equation}
where $\varepsilon = 0.1$ is a nondimensional value representing the variation in the density fluctuation between same-mass aggregates, which is neglected in this study \citep[e.g.,][]{Okuzumi+2011}.
Note that there is a large uncertainty in the estimation of the $\varepsilon$ for CAs; the estimation by \citet{Okuzumi+2011} is only valid for non-compressed fractal aggregates composed of $10$--$10^{6}$ monomers.
For the intermediate region, we assume $\Delta v_{\rm t}^{\rm M}$ such that
\begin{equation}
\Delta v_{\rm t}^{\rm M} = 1.4 {{\rm St}}^{1/2} v_{\rm L}, 
\label{eq13}
\end{equation}
and the relative velocity for high Stokes number aggregates $\Delta v_{\rm t}^{\rm L}$ is
\begin{equation}
\Delta v_{\rm t}^{\rm L} = \left( \frac{2}{1 + {\rm St}} \right)^{1/2} v_{\rm L}.
\end{equation}
Here, we also assume that the turbulence-driven relative velocity $\Delta v_{\rm t}$ is given by the minimum of these three terms:
\begin{equation}
\Delta v_{\rm t} = \min{\left( {\Delta v_{\rm t}^{\rm S}}, {\Delta v_{\rm t}^{\rm M}}, {\Delta v_{\rm t}^{\rm L}} \right)}.
\end{equation}

\subsubsection{Systematic Motions}

Dust aggregates in the gas disk have systematic velocities such as the radial drift velocity, the azimuthal velocity, and the settling velocity.
In this study, we consider the radial drift velocity $v_{\rm r}$ and the azimuthal velocity $v_{\phi}$ given by \citep{Weidenschilling1977}
\begin{equation}
v_{\rm r} = - \frac{2 {\rm St}}{1 + {\rm St}^{2}} \eta v_{\rm K},
\end{equation}
and
\begin{equation}
v_{\phi} = \left( 1 - \frac{1}{1 + {\rm St}^{2}} \right) \eta v_{\rm K},
\end{equation}
where $\eta v_{\rm K}$ is the difference between the Keplerian velocity and the gas rotational velocity.
The Keplerian velocity $v_{\rm{K}}$ is given by $v_{\rm K} = R \Omega_{\rm K}$, and the dimensionless coefficient $\eta$ is given by \citep{Nakagawa+1986}
\begin{equation}
\eta = - \frac{1}{2} {\left( \frac{c_{\rm s}}{v_{\rm K}} \right)}^{2} \frac{\partial \ln{( \rho_{\rm g} {c_{\rm s}}^{2})}}{\partial \ln{R}}.
\end{equation}

\subsection{Density of the Compound Aggregates}

The aerodynamic properties of the aggregates depend on their bulk densities.
In this paper, we assume that all CAs have same mass and density and that the matrix regions of the CAs have a uniform density.
Then, the bulk filling factor of the CAs $\phi_{\rm CA}$ is given by
\begin{equation}
\phi_{\rm CA} = {\left( \frac{\chi}{\phi_{\rm mat}} + \frac{1 - \chi}{1} \right)}^{-1},
\label{eq18}
\end{equation}
where $\phi_{\rm mat}$ is the filling factor of the matrix regions and $\chi$ is the mass fraction of the matrices in a CA.
The bulk density of the CAs $\rho_{\rm CA}$ is
\begin{equation}
\rho_{\rm CA} = \rho_{0} \phi_{\rm CA},
\end{equation}
where $\rho_{0} = 3\ {\rm g}\ {\rm cm}^{-3}$ is the material density of the chondrules and matrix grains.

There are several processes that affect the density of the dust aggregates: the hit-and-stick growth of two colliding aggregates without compaction \citep[e.g.,][]{Meakin1991,Okuzumi+2009}, the dynamical compression via high-speed collisions \citep[e.g.,][]{Paszun+2009,Suyama+2012}, and static compression via the ram pressure of the disk gas and/or via the self-gravity of large planetary bodies \citep[e.g.,][]{Seizinger+2012,Kataoka+2013a}.

Dynamical compression, however, does not work in our calculations, and the CAs are compressed by the static pressure.
Note that \citet{Ormel+2008} assumed that dust rims around chondrules are formed via direct accretion of monomer grains.
Therefore, \citet{Ormel+2008} assumed that the initial filling factor of the matrix regions of CAs is $\phi_{\rm mat} = 0.15$, which is the filling factor obtained by the particle-cluster aggregation.
By contrast, we assume chondrules obtain fluffy dust rims via the accretion of highly porous MAs, and $\phi_{\rm mat}$ is far smaller than $0.15$ at the start of Stage \mbox{I\hspace{-.1em}I\hspace{-.1em}I} in our calculations.
In addition, \citet{Ormel+2008} employs the collisional compression model obtained by \citet{Ormel+2007b}; however, \citet{Wada+2008} revealed that \citet{Ormel+2007b} overestimates the density of fluffy aggregates at the maximum compression.

\citet{Kataoka+2013a} revealed that the static compression of a highly porous aggregate depends on the pressure that works on the aggregate $P$ and the rolling energy $E_{\rm roll}$.
The rolling energy is given by \citep{Dominik+1995}
\begin{equation}
E_{\rm roll} = 6 {\pi}^{2} \gamma a_{0} \xi,
\end{equation}
where $\gamma$ is the surface energy of silicate, $a_{0}$ is the radius of the monomers (i.e., matrix grains), and $\xi$ is the critical displacement for rolling.
We assume that the radius of the matrix grains is $a_{0} = 0.025\ {\mu}{\rm m}$ in this study, which is the typical radius of opaque grains in pristine chondrites such as Acfer 094 and QUE 99177 \citep{Vaccaro+2015}.
We employ the canonical value of the surface energy of silicate, which is $ \gamma = 25\ {\rm erg}\ {\rm cm}^{-2}$ \citep{Kendall+1987}.
The lower limit of $\xi$ is $0.2\ {\rm nm}$, which is equal to the interatomic distance \citep{Dominik+1997}, and the upper limit is the radius of the contact surface area of two matrix grains $a_{\rm cont}$ \citep{Heim+1999}.
\citet{Krijt+2014} showed that $\xi$ is approximately $0.3$--$1\ {\rm nm}$ for micron-sized amorphous silica spheres, and we assume that the critical displacement for rolling is $\xi = 0.3\ \rm{nm}$ for submicron-sized monomers.
Therefore, we can calculate the filling factor of a highly porous dust aggregate using the relation obtained by \citet{Kataoka+2013a}, and the filling factor of matrices in CAs is given by
\begin{equation}
\phi_{\rm mat} = {\left( \frac{{a_{0}}^{3} P}{E_{\rm roll}} \right)}^{1/3}.
\end{equation}
The sources of the pressure $P$ are the ram pressure $P_{\rm gas}$ and the self-gravitational pressure $P_{\rm grav}$ \citep{Kataoka+2013b}.
For CAs whose radii are $a_{\rm CA}$ and masses are $m_{\rm CA}$, the ram pressure of the disk gas is estimated to be $P_{\rm gas} = {(m_{\rm CA} v)} / {(\pi {a_{\rm CA}}^{2} t_{\rm s})}$, whereas the self-gravitational pressure is $P_{\rm grav} = {(G {m_{\rm CA}}^{2})} / {(\pi {a_{\rm CA}}^{4})}$.
We calculate the filling factors and densities assuming that $P = \max{( P_{\rm gas}, P_{\rm grav})}$.

\subsection{Penetration of Chondrules in Matrices}

\begin{figure}[h]
\includegraphics[width=\columnwidth]{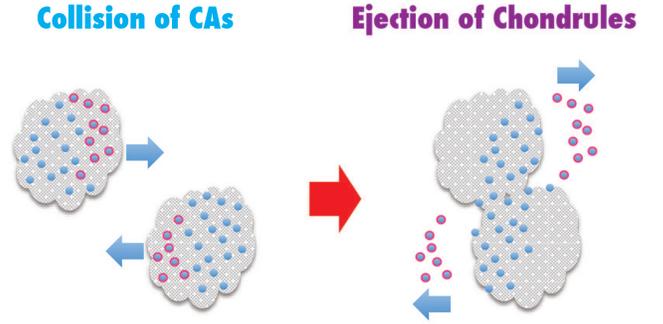}
\caption{Sketch of the outcome of a collision between CAs.
The blue circles represent chondrules, and some of the chondrules, which are rimmed with purple, are ejected when the CAs collide.
}
\label{fig2}
\end{figure}

Chondrules contained in fluffy matrices move in the aggregates when fluffy aggregates collide at large velocities.
We estimate that these CAs cannot retain the chondrules inside their matrices if the stopping length of the chondrules exceeds the radius of the CAs (Figure \ref{fig2}).
Here, we assume that the stopping length is the penetration length for stopping via deceleration forces, $L_{\rm pen}$.

The penetration of chondrules in fluffy matrices is the same process as that for the penetration of spherical projectiles in a silica aerogel \citep{Niimi+2011} and the intrusion of chondrule-analogs into matrix analogs \citep{Machii+2013}.
Therefore, the equation of motion is
\begin{equation}
m_{\rm ch} \frac{{\rm d}v_{\rm pen}}{{\rm d}t} = F_{\rm pen},
\end{equation}
where $m_{\rm ch}$ is the mass of the chondrules.
In this study, we assume that the radius of the chondrules is $a_{\rm ch} = 0.01\ {\rm cm}$, which is a typical size of chondrules in H ordinary chondrites and EH enstatite chondrites \citep{Rubin2000}.
The mass of the chondrules is $m_{\rm ch} = {(4 \pi / 3)} {a_{\rm ch}}^{3} \rho_{0}$.
Note that some other groups (e.g., L and LL ordinary chondrites) contain chondrules whose radii are a few times larger than that which we assume here.

There are two mechanisms for decelerating chondrules in matrices: ``hydrodynamic'' drag for high-speed penetration and ``crushing'' drag for low-speed penetration \citep[e.g.,][]{Niimi+2011}.
In the case of high-speed penetration, the deceleration process is due to the ``hydrodynamic'' drag caused by the ram pressure, and the drag force $F_{\rm pen}$ is given by
\begin{equation}
F_{\rm pen}^{\rm HD} = - \frac{C_{\rm d}}{2} \pi {a_{\rm ch}}^{2} \rho_{\rm mat} {v_{\rm pen}}^{2},
\end{equation}
where $C_{\rm d} = 1$ is the drag coefficient obtained from impact experiments \citep[e.g.,][]{Trucano+1995}.
Conversely, in the case of low-speed penetration, the deceleration process is due to ``crushing'' against the compressive strength of the matrices and the equation of motion is given by
\begin{equation}
F_{\rm pen}^{\rm CR} = - \pi {a_{\rm ch}}^{2} P_{\rm c},
\end{equation}
where $P_{\rm c}$ is the compressive strength of the matrices \citep[e.g.,][]{Dominguez+2004}.
The compressive strength $P_{\rm c}$ has been studied in experiments \citep[e.g.,][]{Blum+2004,Guttler+2009} and in numerical simulations \citep[e.g.,][]{Seizinger+2012,Kataoka+2013a}.
In this study, we calculate the $P_{\rm c}$ of highly porous matrices using the following equation \citep{Kataoka+2013a}:
\begin{equation}
P_{\rm c} = \frac{E_{\rm roll}}{{a_{0}}^{3}} {\phi_{\rm mat}}^{3}.
\end{equation}

Here, we can rewrite the equation of motion as
\begin{equation}
\frac{{\rm d}v_{\rm pen}}{{\rm d}t} = - \max{\left( A {v_{\rm pen}}^{2}, B \right)},
\end{equation}
where $A = {3 C_{\rm d} \phi_{\rm mat}} / {(8 a_{\rm ch})}$ and $B = {3 P_{\rm c}} / {(4 \rho_{0} a_{\rm ch})}$.
Then we can obtain the penetration length $L_{\rm pen}$ as a function of the initial penetration velocity $v_{0}$.
In the case of high-speed penetration ($v_{0} \geq \sqrt{B / A}$), $L_{\rm pen}$ is given by
\begin{equation}
L_{\rm pen} = \frac{1}{A} {\rm ln} {\left( v_{0} \sqrt{A / B} \right)} + \frac{1}{2 A},
\label{eq26}
\end{equation}
and in the case of low-speed penetration ($v_{0} < \sqrt{B / A}$), $L_{\rm pen}$ is given by
\begin{equation}
L_{\rm pen} = \frac{{v_{0}}^{2}}{2 B}.
\label{eq27}
\end{equation}

If two colliding CAs have the same mass and the matrices of the two colliding aggregates stop immediately after the collision, then the initial penetration velocity $v_{0}$ can be evaluated as ${\Delta v} / 2$.
However, the motion of the matrices after the collision is not yet understood.
We examine two cases in this study: $v_{0} = {\Delta v} / 2$ and $v_{0} = {\Delta v} / 8$.
Note that, if the mass ratio of the projectile to the target is small, then $v_{0}$ of the target decreases while $v_{0}$ of the impactor increases when considering the motions relative to the center of mass.

\subsection{Critical Velocity for Collisional Growth}

We also estimate the critical velocity for the collisional growth of aggregates ${\Delta v}_{\rm crit}$ from the scaling relation obtained by \citet{Wada+2009}.
The impact energy of two colliding aggregates $E_{\rm imp}$ is $E_{\rm imp} = m {(\Delta v)}^{2} / 4$, and the critical impact energy for growth $E_{\rm imp, crit}$ is $E_{\rm imp, crit} = 30 N_{\rm total} E_{\rm break}$, where $N_{\rm total}$ is the total number of monomers contained in the colliding aggregates, and $E_{\rm break}$ is the energy for breaking a single contact between two monomers \citep{Wada+2009}.
The energy for breaking a monomer-monomer contact $E_{\rm break}$ is given by \citet{Wada+2007}.
Clearly, the total binding energy of a CA is dominated by the contacts between matrix grains.
Therefore, the total number of monomers in the two colliding CAs is $N_{\rm total} = 2 \chi m_{\rm CA} / m_{0}$, where $m_{0}= {(4 \pi / 3)} {a_{0}^{3}} \rho_{0}$ is the mass of monomers, and the critical velocity for the collisional growth of CAs ${\Delta v}_{\rm crit}$ is given by
\begin{equation}
{\Delta v}_{\rm crit} = \sqrt{\frac{240 \chi E_{\rm break}}{m_{0}}} = 5.5 \times 10^{2}\ {\rm cm}\ {\rm s}^{-1}.
\end{equation}
Note, however, that the critical collision velocity of dust aggregates is still open to argument \citep[e.g.,][]{Tanaka+2012,Krijt+2013}.
Hence we do not consider fragmentation of CAs in this study.

\section{Results}
\label{results}

\begin{figure}[h]
\includegraphics[width=\columnwidth]{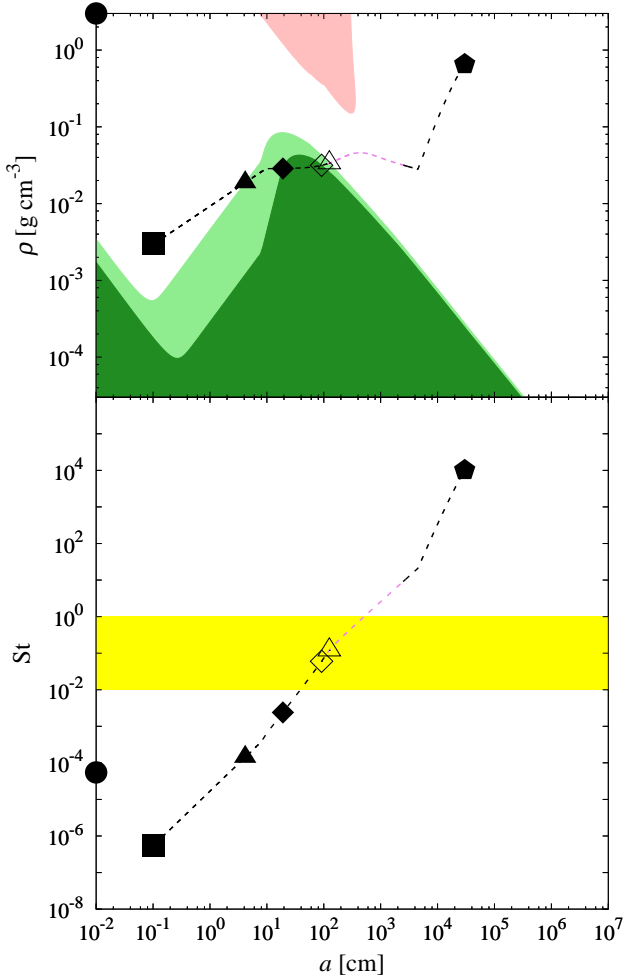}
\caption{
The evolutionary track of the CAs (the black and violet dashed lines), the radial drift barrier (the pink region), and the ejection barrier (the forest- and light-green regions), for the case of $\dot{M} = 10^{-7}\ M_{\odot}\ {\rm yr}^{-1}$, $R = 1\ {\rm au}$, and $Z = 0.0043$.
The black lines represent where $\Delta v < {\Delta v}_{\rm crit}$ and the violet lines represent where $\Delta v > {\Delta v}_{\rm crit}$.
The upper panel shows that the density evolution track of CAs is determined by static compression.
We show two barriers in radius-density space.
The lower panel represents the change in the Stokes number as a function of the radius.
It also indicates the appropriate area for driving the streaming instability in a disk with a moderate dust-to-total disk mass ratio, $10^{-2} \lesssim {\rm St}_{\rm CA} \lesssim 1$ (the yellow region).
The circles indicate the values of single chondrules, the filled triangles and diamonds indicate the onset of critical ejection ($L_{\rm pen} = a_{\rm CA}$) while the open triangles and diamonds represent the ends.
Triangles correspond to the severe estimate ($v_{0} = {\Delta v} / 2$) and diamonds correspond to the lenient estimate ($v_{0} = {\Delta v} / 8$).
The pentagonal markers indicate the onset of runaway growth (${\Delta v} = \sqrt{2 G m_{\rm CA} / a_{\rm CA}}$).
}
\label{fig3}
\end{figure}

First, we calculate the pathway of the dust aggregate growth in radius-density space and investigate whether the growth of the CAs is sufficiently rapid to avoid non-negligible radial drift.
Here, we neglect bouncing, erosion, and fragmentation for simplicity; therefore, the timescale of collisional growth $t_{\rm grow}$ is defined as $t_{\rm grow} = m / {(\pi a^{2} \rho_{\rm d} {\Delta v})}$, where $\rho_{\rm d} = \Sigma_{\rm d} / {(\sqrt{2 \pi} h_{\rm d})}$ is the midplane dust density and $h_{\rm d}$ is the dust scale height.
The dust scale height is given by \citep{Youdin+2007}
\begin{equation}
h_{\rm d} = h_{\rm g} {\left( 1 + \frac{\rm St}{\alpha_{\rm t}} \frac{1 + 2 {\rm St}}{1 + {\rm St}} \right)}^{- 1/2}.
\label{eq28}
\end{equation}
The timescale of radial drift $t_{\rm drift}$ is defined as the orbital radius divided by the radial drift velocity, $t_{\rm drift} = R / v_{\rm r}$.
Aggregates grow without significant radial drift if the condition $t_{\rm grow} < {(1 / 30)} t_{\rm drift}$ is satisfied \citep{Okuzumi+2012}.
Therefore, the radial drift barrier (the pink region) and the growth pathway of the CAs (the black and violet dashed lines) can be drawn as shown in Figure \ref{fig3}.
The black lines represent where the collision velocity $\Delta v$ is below the critical velocity ${\Delta v}_{\rm crit}$, while the violet lines represent where $\Delta v > {\Delta v}_{\rm crit}$.
We also calculated the the penetration length of the chondrules in CAs $L_{\rm pen}$ as a function of $a_{\rm CA}$ and $\rho_{\rm CA}$.
In Figure \ref{fig3}, we show the ``ejection barrier,'' which represents the region where the penetration length $L_{\rm pen}$ exceeds the radius of the CAs $a_{\rm CA}$.
The forest-green region represents the ejection barrier obtained by assuming that the initial penetration velocity of $v_{0} = {\Delta v} / 2$, while the light-green region represents the case where $v_{0} = {\Delta v} / 8$.

The upper panel of Figure \ref{fig3} illustrates that the pathway of the CAs overcomes the radial drift barrier but fails to avoid the ejection barrier.
In this case, we assume $\dot{M} = 10^{-7}\ M_{\odot}\ {\rm yr}^{-1}$ \citep[which is comparable to the accretion rate of the disk around HL Tau;][]{Beck+2010}, $R = 1\ {\rm au}$ (which is the distance from the Sun to the Earth), and $Z = 0.0043$ \citep[e.g.,][]{Miyake+1993} in this case.
The radius and the density of the chondrules are represented by the circular marker, and the radius and the density of the CAs, which contain one chondrule, are represented by the square marker.
Thus the square marker is equivalent to the start of Stage \mbox{I\hspace{-.1em}I\hspace{-.1em}I}.
If we do not consider the ejection of chondrules from the fluffy matrices, the CAs can grow into planetesimal without radial drift and the pentagonal marker represents the onset of runaway growth \citep[${\Delta v} = \sqrt{2 G m_{\rm CA} / a_{\rm CA}}$;][]{Wetherill+1989,Kobayashi+2016}.
Note that the onset of runaway growth might be affected by turbulence-induced density fluctuations \citep{Okuzumi+2013,Ormel+2013} and/or the gravitational instability of the porous planetesimal disk \citep{Michikoshi+2016,Michikoshi+2017}.
Therefore, we do not discuss the onset of runaway growth in this study but focus on the growth of CAs whose Stokes numbers are less than unity.

The filled diamond indicates the onset of critical ejection, $L_{\rm pen} = a_{\rm CA}$, for the severe estimate ($v_{0} = {\Delta v} / 2$), and the open diamond indicates the end of critical ejection for the severe estimate.
Similarly, the filled and open triangles indicate the onset and end of critical ejection for the lenient estimate ($v_{0} = {\Delta v} / 8$).
Our results show that, if the density of the CAs is determined by static compression, then CAs cannot grow into planetesimals without losing chondrules, even if we evaluate the penetration length using a lenient assumption.
Our calculations also reveal that CAs face the ejection barrier before they grow into meter-sized bodies and face the fragmentation barrier.

In addition, the recapture of chondrules by chondrule-free aggregates is not sufficient to explain the formation of chondrite parent bodies.
The lower panel in Figure \ref{fig3} shows that the Stokes numbers of the chondrules and the CAs.
The Stokes number of the chondrules is ${\rm St}_{\rm ch} = 5.5 \times 10^{-5}$, and the Stokes number of the CAs when they reach the end of critical ejection is ${\rm St}_{\rm CA} = 0.12$ for the severe estimate and ${\rm St}_{\rm CA} = 6.0 \times 10^{-2}$ for the lenient estimate.
According to Equation (\ref{eq28}), the scale height of the CAs is $h_{\rm CA} \sim {(1 / 30)} h_{\rm g}$ while the scale height of the chondrules is $h_{\rm ch} \sim h_{\rm g}$.
The collision velocity of two CAs is $1.4 {{\rm St}_{\rm CA}}^{1/2} v_{\rm L}$ (Equation \ref{eq13}) and the relative velocity between the colliding chondrule to the CA is approximately $1.7 {{\rm St}_{\rm CA}}^{1/2} v_{\rm L}$ (Equation \ref{eq9}).
This is because the Stokes number of the chondrules ${\rm St}_{\rm ch}$ is far smaller than the Stokes number of the CAs ${\rm St}_{\rm CA}$.
Therefore, the timescale for chondrules to accrete onto CAs $t_{{\rm ch} \to {\rm CA}}$ is dozens of times longer than the timescale of collisions between two CAs $t_{{\rm CA} \to {\rm CA}}$.
This result indicates that CAs will have already grown much more than $2^{10}$ times in mass and more than ten times in radii, before the chondrules accrete onto the CAs.
Conversely, the penetration length $L_{\rm pen}$ is given in Equation (\ref{eq26}) and $L_{\rm pen} \sim {1 / A}$ hardly changes even when the CAs grow ten times in radii.
Strictly speaking, the initial penetration velocity $v_{0}$ is not ${\Delta v} / 2$ but approximately $v$ for the case of recapture: however, this would not significantly affect the penetration length for the reason described above.
Therefore, taking into consideration that the radius of the CAs is on the order of $100\ {\rm cm}$ at the end of critical ejection, the chondrite parent bodies would consist of rocks with meter-sized chondrule-free matrices.
This is, however, inconsistent with observational evidence.
One of the most important pieces of evidence is that the most numerous type of meteorites is chondrites \citep{Scott2007}, and the typical size of the recovered chondrites is $10\ {\rm cm}$.
This fact excludes the possibility of the recapture of chondrules by meter-sized fluffy aggregates.

\begin{figure*}
\vspace{-5truemm}
\gridline{\fig{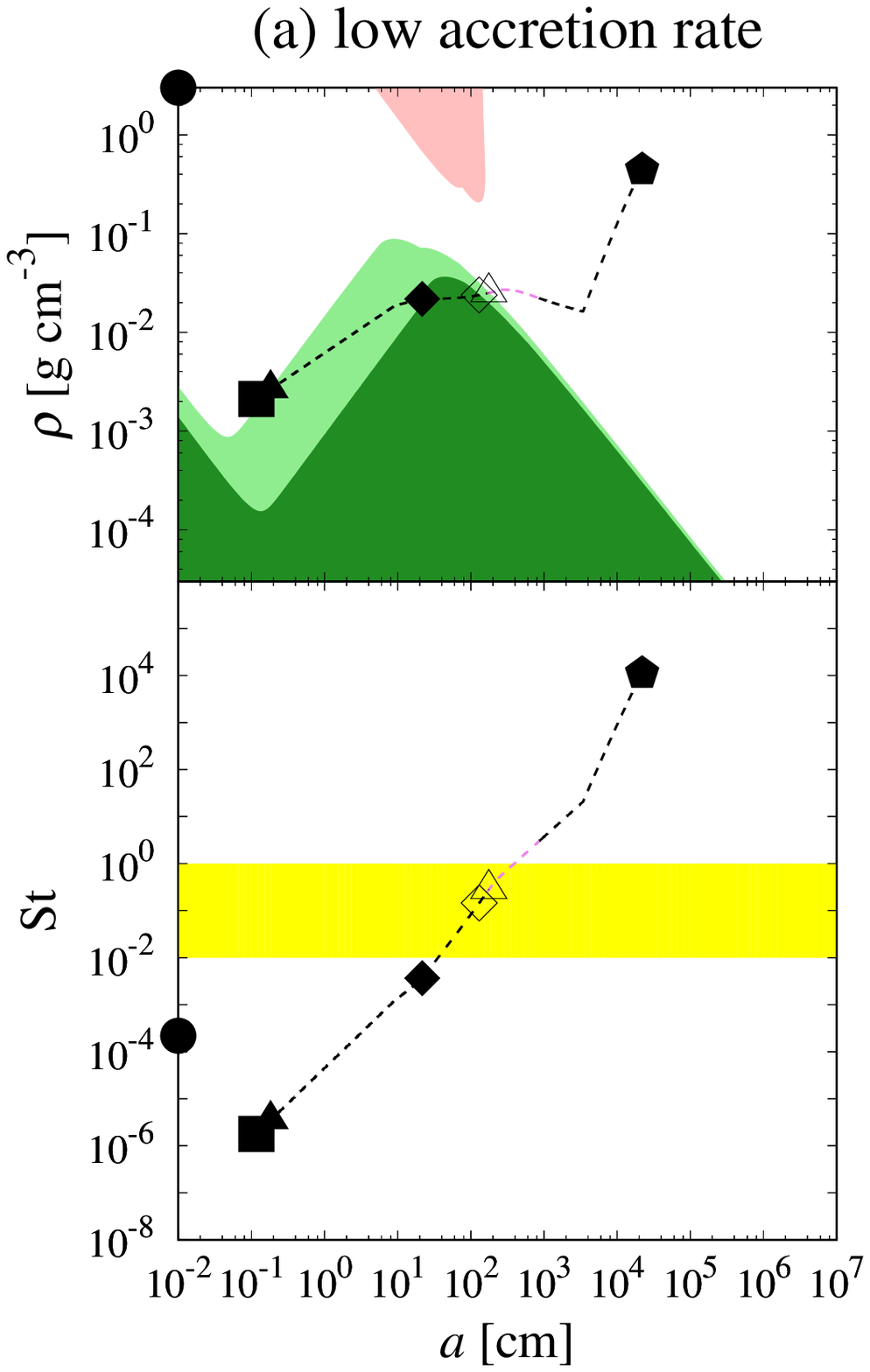}{0.33\textwidth}{}
          \fig{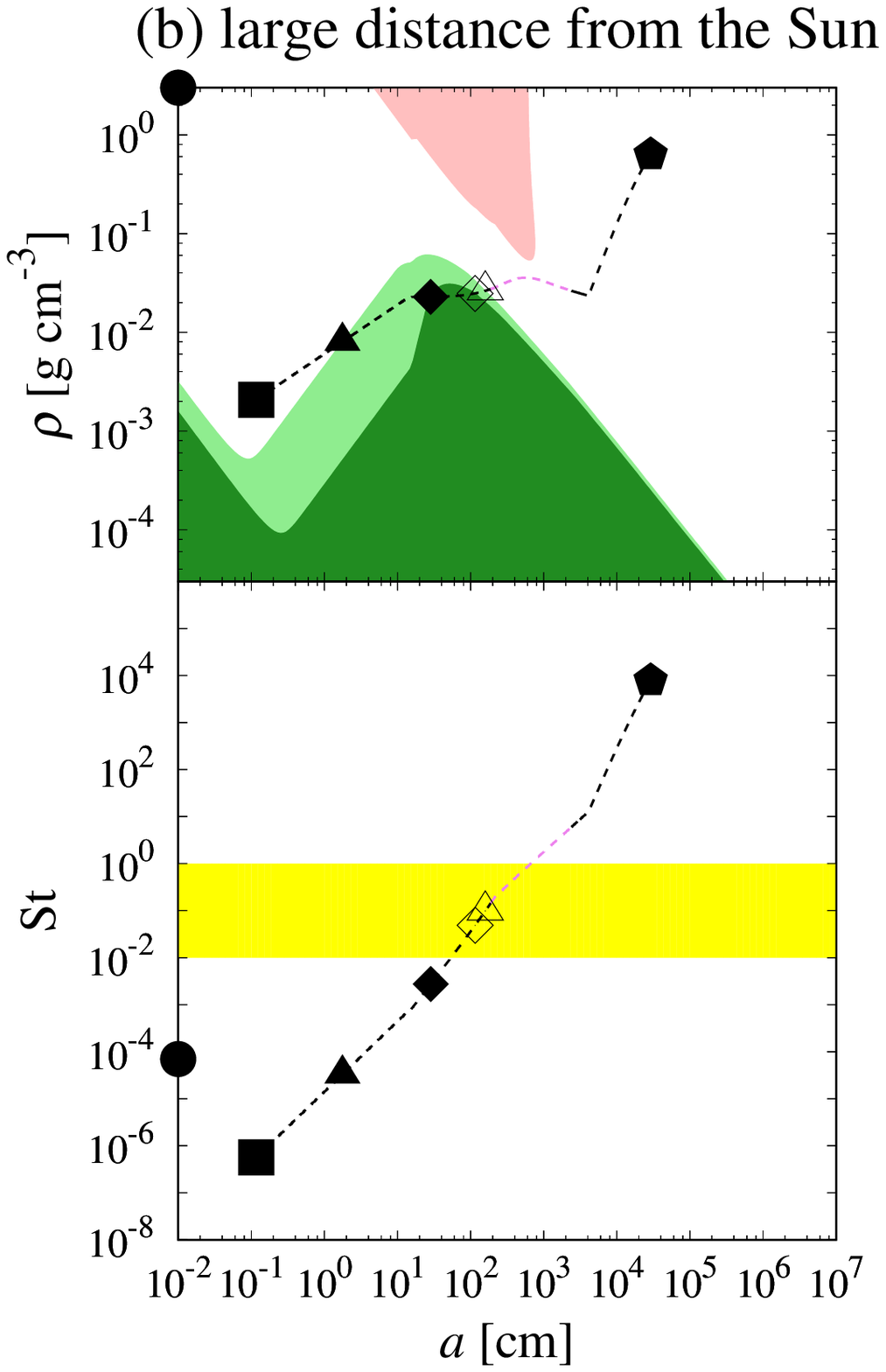}{0.33\textwidth}{}
          \fig{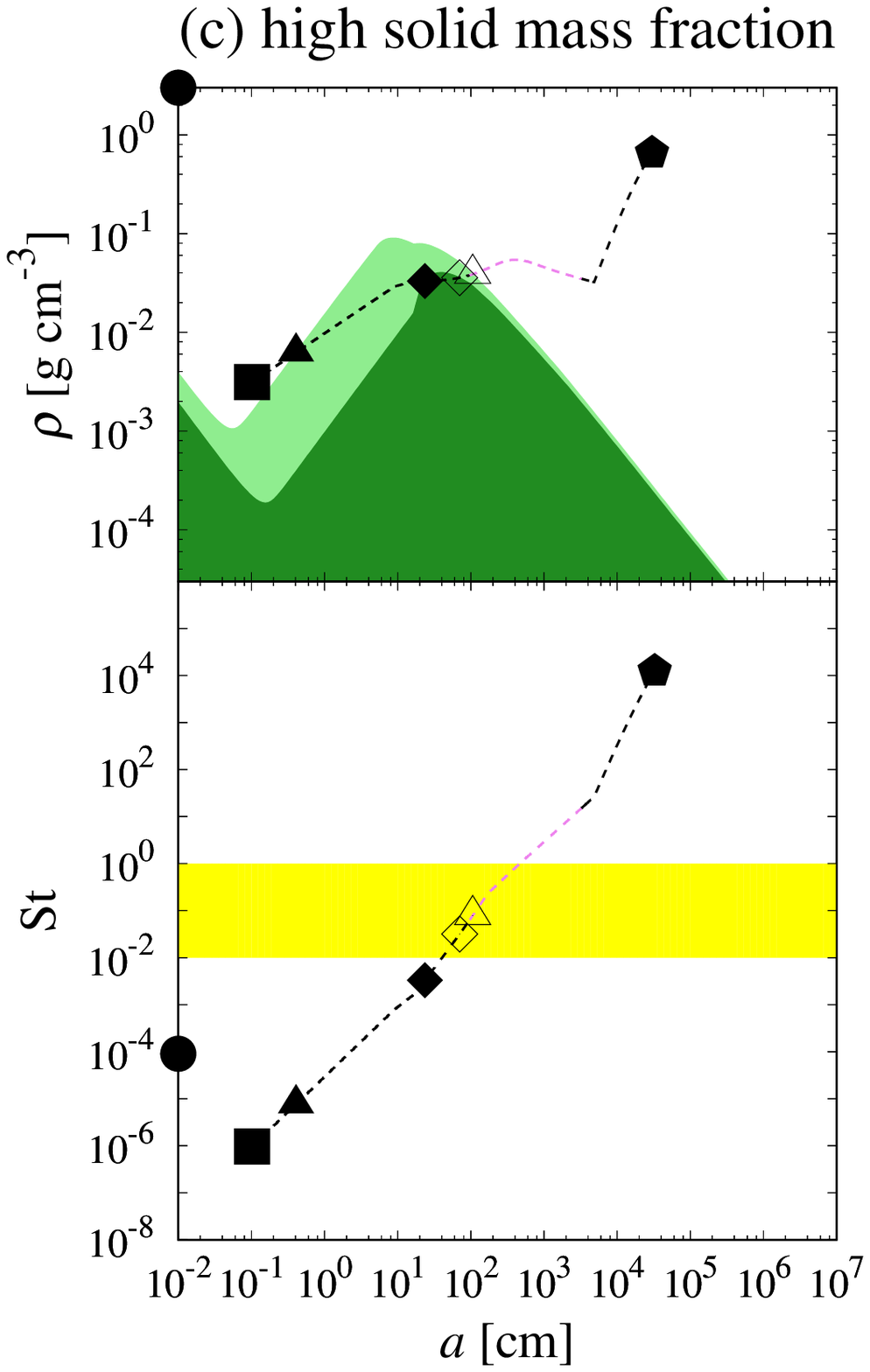}{0.33\textwidth}{}}
\vspace{-10truemm}
\caption{Same as Figure \ref{fig3} but for different disk parameters:
(a) the case of $\dot{M} = 10^{-8}\ M_{\odot}\ {\rm yr}^{-1}$, $R = 1\ {\rm au}$, and $Z = 0.0043$;
(b) the case of $\dot{M} = 10^{-7}\ M_{\odot}\ {\rm yr}^{-1}$, $R = 1.5\ {\rm au}$, and $Z = 0.0043$;
and (c) the case of $\dot{M} = 10^{-7}\ M_{\odot}\ {\rm yr}^{-1}$, $R = 1\ {\rm au}$, and $Z = 0.043$.
}
\label{fig4}
\end{figure*}

We show the results for other disk parameters in Figure \ref{fig4}.
Figure \ref{fig4}a shows the result for a case with $\dot{M} = 10^{-8}\ M_{\odot}\ {\rm yr}^{-1}$, $R = 1\ {\rm au}$, and $Z = 0.0043$.
This is a low-accretion-rate case compared to the case in Figure \ref{fig3}.
The result shows that CAs cannot retain chondrules even in this case, and that the onset of critical ejection strongly depends on the estimate of the initial penetration velocity $v_{0}$.
Conversely, the end of critical ejection depends little on the initial penetration velocity $v_{0}$.
This is because the onset is determined by Equation (\ref{eq27}), so the penetration length depends on ${v_{0}}^{2}$, while the end is determined by Equation (\ref{eq26}), so the dependence of $L_{\rm pen}$ on $v_{0}$ is logarithmic.

Figure \ref{fig4}b shows the result for the case of $\dot{M} = 10^{-7}\ M_{\odot}\ {\rm yr}^{-1}$, $R = 1.5\ {\rm au}$, and $Z = 0.0043$.
The difference between this case and the case in Figure \ref{fig3} is the distance from the Sun.
The radial drift barrier is sensitive to the radial distance from the Sun, so CAs can grow into planetesimals only within $R \lesssim 1.5\ {\rm au}$ even if we do not consider the ejection and fragmentation problems.
Similarly, Figure \ref{fig4}c shows the result for the case of $\dot{M} = 10^{-7}\ M_{\odot}\ {\rm yr}^{-1}$, $R = 1\ {\rm au}$, and $Z = 0.043$ and the difference between this case and the case in Figure \ref{fig3} is the mass fraction of dust.
The radial drift barrier is also sensitive to the dust mass fraction $Z$, and there is no radial drift barrier if $Z$ is approximately ten times higher than the canonical silicate abundance inferred from the solar metallicity \citep[$Z = 0.0043$;][]{Miyake+1993}.
Nevertheless, CAs might lose chondrules before their Stokes number reaches approximately $10^{-2}$.
The possibility of the recapture of chondrules by grown fluffy aggregates is however denied in either case for the same reason as that in the case of Figure \ref{fig3}.

\section{Analytic Estimates}
\label{analytic}

From the results shown in Figures \ref{fig3} and \ref{fig4}, we can divide the ejection barrier into four regions; the Brownian-motion region, the Epstein-drag region, the Stokes-drag region, and the hydrodynamical region.
In this section, we derive analytical formulae for the Epstein-drag region, the Stokes-drag region, and the hydrodynamic region.

\subsection{Epstein-Drag Region}
\label{epstein}

First, we define the ``Epstein-drag region'' as the region in which the CA affects the Epstein drag and the chondrules are decelerated by the compressive stress of the matrices.
In the upper panel in Figure \ref{fig3}, this region appears within the range between approximately $0.3\ {\rm cm}$ and $8\ {\rm cm}$ for the lenient case and between approximately $0.1\ {\rm cm}$ and $8\ {\rm cm}$ for the severe case.
From our calculation, the collision velocity is given by Equation (\ref{eq12}) and we can rewrite Equation (\ref{eq27}) as
\begin{equation}
L_{\rm pen} \simeq \frac{\pi^{2}}{6} {\left( \frac{v_{0}}{\Delta v} \right)}^{2} \frac{\alpha_{\rm t} \varepsilon^{2} {{\rm Re}_{\rm t}}^{1/2} {c_{\rm s}}^{2}}{{\Sigma_{\rm g}}^{2}} \frac{{\rho_{0}}^{3} {a_{\rm CA}}^{2}}{\chi^{3} \phi_{\rm CA}} {\left( \frac{E_{\rm roll}}{{a_{0}}^{3}} \right)}^{- 1} a_{\rm ch}.
\end{equation}
Here, we use the fact that the volume of a fluffy CA is dominated by its matrices; therefore, the filling factor of the matrices $\phi_{\rm mat}$ is nearly equal to $\chi \phi_{\rm CA}$ according to Equation (\ref{eq18}).
Then, the critical bulk density for keeping chondrules in their matrices, obtained from the requirement $L_{\rm pen} \lesssim a_{\rm CA}$, is given by
\begin{equation}
\rho_{\rm CA} \gtrsim \frac{\pi^{2}}{6} {\left( \frac{v_{0}}{\Delta v} \right)}^{2} \frac{\alpha_{\rm t} \varepsilon^{2} {{\rm Re}_{\rm t}}^{1/2} {c_{\rm s}}^{2}}{{\Sigma_{\rm g}}^{2}} \frac{{\rho_{0}}^{4} a_{\rm ch}}{\chi^{3}} {\left( \frac{E_{\rm roll}}{{a_{0}}^{3}} \right)}^{- 1} a_{\rm CA}.
\end{equation}

\subsection{Stokes-Drag Region}

As in Section \ref{epstein}, we introduce the ``Stokes-drag region'' as the region in which the CA affects the Stokes drag and the chondrules are decelerated by the compressive stress of the matrices.
In the upper panel in Figure \ref{fig3}, this region appears within the range from approximately $8\ {\rm cm}$ to near the summit of the ejection barrier for both the lenient and severe cases.
The collision velocity is given by Equation (\ref{eq12}), the penetration length is
\begin{equation}
L_{\rm pen} \simeq \frac{16 \pi}{243} {\left( \frac{v_{0}}{\Delta v} \right)}^{2} \frac{\varepsilon^{2} {{\rm Re}_{\rm t}}^{5/2} {\Omega_{\rm K}}^{2}}{\alpha_{\rm t} {\Sigma_{\rm g}}^{2}} \frac{{\rho_{0}}^{3} {a_{\rm CA}}^{4}}{\chi^{3} \phi_{\rm CA}} {\left( \frac{E_{\rm roll}}{{a_{0}}^{3}} \right)}^{- 1} a_{\rm ch},
\end{equation}
and the critical bulk density for retaining the chondrules in their matrices is given by
\begin{equation}
\rho_{\rm CA} \gtrsim \frac{16 \pi}{243} {\left( \frac{v_{0}}{\Delta v} \right)}^{2} \frac{\varepsilon^{2} {{\rm Re}_{\rm t}}^{5/2} {\Omega_{\rm K}}^{2}}{\alpha_{\rm t} {\Sigma_{\rm g}}^{2}} \frac{{\rho_{0}}^{4} a_{\rm ch}}{\chi^{3}} {\left( \frac{E_{\rm roll}}{{a_{0}}^{3}} \right)}^{- 1} {a_{\rm CA}}^{3}.
\end{equation}
The Stokes-drag region does not appear in the severe cases in Figures \ref{fig4}a, \ref{fig4}b, and \ref{fig4}c.
This is because the initial velocity of penetration $v_{0}$ already exceeds $\sqrt{B / A}$ before the CAs enter the Stokes-drag region when they grow along the critical ejection lines in radius-density space.

Note that the relative velocity $\Delta v$ changes $\Delta v_{\rm t}^{\rm S} = \varepsilon {{\rm Re}_{\rm t}}^{1/4} {\rm St} v_{\rm L}$ (Equation \ref{eq12}) into $\Delta v_{\rm t}^{\rm M} = 1.4 {{\rm St}}^{1/2} v_{\rm L}$ (Equation \ref{eq13}) when the Stokes number reaches ${\rm St} = {(\varepsilon / 1.4)}^{-2} {{\rm Re}_{\rm t}}^{- 1/2}$, and the critical Stokes number for $\varepsilon = 0.1$ is on the order of $10^{-2}$; however, the critical Stokes number strongly depends on the uncertain parameter $\varepsilon$.
If $\varepsilon$ is close to unity, the relative velocity of the CAs is given by Equation (\ref{eq13}), and the penetration length is given by
\begin{equation}
L_{\rm pen} \simeq 0.73 {\left( \frac{v_{0}}{\Delta v} \right)}^{2} \frac{{\rm Re}_{\rm t} c_{\rm s} \Omega_{\rm K}}{\Sigma_{\rm g}} \frac{{\rho_{0}}^{2} {a_{\rm CA}}^{2}}{\chi^{3} {\phi_{\rm CA}}^{2}} {\left( \frac{E_{\rm roll}}{{a_{0}}^{3}} \right)}^{- 1} a_{\rm ch}.
\end{equation}
In this case, the critical bulk density is
\begin{equation}
\rho_{\rm CA} \gtrsim 0.85 \frac{v_{0}}{\Delta v} \sqrt{\frac{{\rm Re}_{\rm t} c_{\rm s} \Omega_{\rm K}}{\Sigma_{\rm g}}} \frac{{\rho_{0}}^{2} {a_{\rm ch}}^{1/2}}{\chi^{3/2}} {\left( \frac{E_{\rm roll}}{{a_{0}}^{3}} \right)}^{- 1/2} {a_{\rm CA}}^{1/2}.
\end{equation}

\subsection{Hydrodynamic Region}

Finally, we define the ``hydrodynamic region'' as the region in which chondrules in a CA are decelerated by the ram pressure caused by high-speed penetration.
The penetration length is given by Equation (\ref{eq26}), and we assume that the term $\ln{(v_{0} \sqrt{A / B})}$ is on the order of unity; then, we can rewrite Equation (\ref{eq26}) as
\begin{equation}
L_{\rm pen} \sim \frac{8 \chi}{3 \phi_{\rm CA}} a_{\rm ch}.
\end{equation}
The critical bulk density for retaining chondrules is
\begin{equation}
\rho_{\rm CA} \gtrsim \frac{8 \chi \rho_{0} a_{\rm ch}}{3} {a_{\rm CA}}^{- 1}.
\end{equation}
The critical bulk density depends little on the properties of matrix grains in the hydrodynamic region, while the properties of matrix grains (i.e., $E_{\rm roll} / {a_{0}}^{3}$) strongly affect the critical bulk density in the Epstein-drag and Stokes-drag regions.

In all regions, the penetration length is proportional to the radius of the chondrules.
We assume $a_{\rm ch} = 0.01\ {\rm cm}$ in this study, however, the sizes of chondrules vary between different types of chondrites.
The mean radius of chondrules in H ordinary chondrites and EH enstatite chondrites is approximately $0.01\ {\rm cm}$ whereas the chondrules in L and LL ordinary chondrites and EL ordinary chondrites are approximately $0.03\ {\rm cm}$ in radii \citep[][and references therein]{Rubin2000}.
By contrast, the spread of sizes in individual groups is relatively small.
Approximately 70\% of chondrules in Qingzhen, Kota-Kota, and Allan Hills A77156 EH3 chondrites have similar radii within a factor of two \citep{Rubin+1987}.
It might be possible to derive some constraints on the accretion environment of CAs from the differences in chondrule size in the different groups of chondrites.
We will discuss this topic in the future.

\section{Discussion} \label{discussions}

\subsection{The Complementarity of Chondrules and Matrices}

Recent high-precision measurements of chondrules, matrices, and bulk chondrites reveal that chondrites have chemical \citep[e.g.,][]{Palme+2015,Ebel+2016} and isotopic \citep[e.g.,][]{Budde+2016a,Budde+2016b} complementarities.
These complementarities are therefore a strong constraint, indicating that the chondrules and the matrices must have formed from a single reservoir and that after their formation, neither the chondrules nor the matrix grains were lost.
The simplest way to explain the chondrule-matrix complementarity is that the chondrules and the matrix grains were formed in the same heating events and that some parts of the matrix grains are condensates of evaporated dust.
\citet{Miura+2010} showed that a shock wave heating model for chondrule formation can predict the evaporation and formation of fine dust grains via chondrule formation; in addition, \citet{Miura+2005} revealed that the minimum size of chondrules that avoids the complete evaporation is consistent with the observations \citep[e.g.,][]{Eisenhour1996,Nelson+2002}.

There are many models for chondrule formation, and these models are classified into two groups, that is, impact origin models \citep[e.g.,][]{Sanders+2012,Dullemond+2014,Wakita+2017} and nebular origin models \citep[e.g.,][]{Muranushi2010,McNally+2013,Boley+2013}.
To determine the formation process(es) of chondrules, the key constraints are the heating and cooling timescales.
According to \citet{Tachibana+2005}, the chondrule-forming events must be flash heating events, i.e., a heating rate of $10^{4}$--$10^{6}\ {\rm K}\ {\rm h}^{-1}$ would be required to avoid the isotopic fractionation of sulfur in chondrule precursors.
In addition, although there are huge uncertainties, several studies suggest that the cooling rate of chondrules is as large as $10^{5}$--$10^{6}\ {\rm K}\ {\rm h}^{-1}$ from observations of iron-magnesium and oxygen isotopic exchange \citep[e.g.,][]{Yurimoto+2002} and the crystal growth of olivine phenocrysts \citep[e.g.,][]{Wasson+2003,Miura+2014}.
These estimates are larger by several orders of magnitude than the cooling rates of $10$--$10^{3}\ {\rm K}\ {\rm h}^{-1}$ obtained from furnace-faced experiments \citep[][and references therein]{Desch+2012}.

Even though it is true that furnace-faced experiments have succeeded in reproducing some textural features of chondrules \citep[e.g.,][]{Hewins+2004,Tsuchiyama+2004}, the solidification processes might be affected by the contact with a furnace wall \citep{Nagashima+2006}.
Furthermore, rapid cooling of levitating supercooled precursors also succeeds in reproducing some textures of chondrules \citep[e.g.,][]{Srivastave+2010}.
The supercooling of chondrule melts is preferable to explain the textural features of compound chondrules, which might form via the collision of chondrule precursors \citep{Arakawa+2016a}.
The rapid cooling of the chondrules indicates that the evaporated dust would also be cooled rapidly, and the cooling rate of the dust vapor affects the size and the morphology of the condensates \citep[e.g.,][]{Miura+2010}.

\subsection{Influence of the Monomer Properties}

Understanding of the physical properties of the matrix grains is important when considering the growth process of CAs.
As shown in Section \ref{analytic}, the penetration length $L_{\rm pen}$ is inversely proportional to $E_{\rm roll} / {a_{0}}^{3}$, and the rolling energy of two sticking matrix grains $E_{\rm roll}$ is $E_{\rm roll} = 6 \pi^{2} \gamma a_{0} \xi$.
Therefore, we can obtain the relationship between the surface energy, the radius, and the penetration length, $L_{\rm pen} \propto \gamma^{-1} {a_{0}}^{2}$, and large surface energies and small grain sizes are preferable when the bulk density is fixed.
We append that if the bulk density of the CAs is determined by static compression, $\rho_{\rm CA}$ also depends on $E_{\rm roll} / {a_{0}^{3}}$; however, the dependence is weak \citep[e.g.,][]{Kataoka+2013b,Arakawa+2016b}.

Here we mention the fact that the exact value of $E_{\rm roll}$ might differ from $6 \pi^{2} \gamma a_{0} \xi$.
This is because the equation $E_{\rm roll} = 6 \pi^{2} \gamma a_{0} \xi$ obtained by \citet{Dominik+1995} is based on the so-called JKR contact model \citep{Johnson+1971}, and the JKR model is valid only for large values of the Tabor parameter $\mu$ \citep{Tabor1977}:
\begin{equation}
\mu = {\left( \frac{2 a_{0} \gamma^{2} {(1 - \nu)}^{2}}{Y^{2} {z_{0}}^{3}} \right)}^{1/3} = 0.11\ {\left( \frac{a_{0}}{0.025\ {\mu}{\rm m}} \right)}^{1/3},
\end{equation}
where $\nu = 0.17$ is the Poisson's ratio, $Y = 5.4 \times 10^{11}\ {\rm dyn}\ {\rm cm}^{-2}$ is the Young's modulus, and $z_{0} = 0.2\ {\rm nm}$ is the interatomic distance.
The JKR model can be used only for $\mu \gtrsim 5$ \citep{Johnson+1997}, and thus other contact models should be used for submicron-sized silicate spheres.
In addition, the dependence on the sphere radius is weak, thus we should not adopt the JKR theory not only for submicron-sized spheres but also micron- or millimeter-sized silicate grains.
Even if we consider chondrule-sized grains, the Tabor parameter is still small; the Tabor parameter is $\mu = 1.7$ when the sphere radius is $a_{0} = 0.01\ {\rm cm}$.
The Maugis-Dugdale model \citep{Maugis1992} might be better than the JKR model where the Tabor parameter is $0.1 \lesssim \mu \lesssim 5$, and several contact models such as the so-called DMT model \citep{Derjaguin+1975} and the semi-rigid sphere model \citep{Greenwood2007} are suggested for $\mu \lesssim 0.1$.
However, the maximum force needed to separate the two contact spheres $F_{\rm c}$ is hardly dependent on the selection of contact models: $F_{\rm c} = {(3/2)} \pi \gamma a_{0}$ for the JKR model and $F_{\rm c} =  2 \pi \gamma a_{0}$ for the DMT model.
Additionally, the rolling energy $E_{\rm roll}$ is proportional to $F_{\rm c} \xi$ (e.g., $E_{\rm roll} = 4 \pi F_{\rm c} \xi$ for the JKR model), therefore the rolling energy might be also subequal among contact models.
Either way, future studies of the rolling resistance between two stiff spheres are necessary.

The morphology of the matrix grains also influences both the penetration process and the density evolution because the irregular shape of the monomers affects the rolling energy.
In addition, the irregular shape affects not only the density and the penetration process but also the critical velocity for collisional growth \citep{Poppe+2000}.
It has been suggested that at least some parts of the matrix grains are condensates of evaporated dust and that the size of the condensate ranges from nanometer to micrometer \citep[e.g.,][]{Yamamoto+1977,Miura+2010} depending on the cooling rate and the mass density of the dust vapor.
\citet{Ishizuka+2015} experimentally revealed that there are two types of nanometer-sized silicate condensates obtained from the re-condensation of magnesium and silicon monoxide vapor, that is, spherical forsterite particles and cubic MgO particles.
However, there is no reliable contact theory that can be applied to non-spherical monomers.
Therefore, improvements in the contact mechanical theory for non-spherical monomers are also necessary.

\subsection{Dust Rims around the Chondrules}

In this study, we assume the uniform density of the matrix regions throughout the CA for simplicity.
However, the gradient and fluctuation of the density of the matrix regions of a CA might exist to some extent.
Considering the accretion process in Stage \mbox{I\hspace{-.1em}I}, the initial growth of chondrules/CAs is driven by the accretion of small and dense MAs.
If the dense dust rims around chondrules can be maintained when a CA collide at a large velocity, dust-rimmed chondrules in the fluffy CA are subjected to a large drag force $F_{\rm pen}$ and the penetration length $L_{\rm pen}$ of dust-rimmed chondrules might be smaller than the penetration length of naked chondrules.
Although it is still unclear whether chondrules can maintain the dense dust rims or not, we should consider the formation and influence of the dust rims around chondrules.

\subsection{Deceleration by Collisions of Chondrules}

In this paper, we only consider the drag force for penetration $F_{\rm pen}$ as a decelerating mechanism.
Even though the narrow size distribution of chondrules \citep[e.g.,][]{Nelson+2002} might prevent chondrules from colliding unless there is a large variation in the initial velocity of penetration, chondrules in a large CA are likely affected by chondrule collisions when they move in the CA.
In addition, the re-capture of chondrules by large CAs is likely affected by this process because there is a large velocity difference between intruding chondrules and chondrules contained in CAs.
We will examine this effect in the future.

\subsection{Alternative Scenarios}

As shown in Section \ref{results}, CAs would lose their chondrules if the bulk density of these aggregates were determined by static compression processes.
Therefore, we need to consider alternative scenarios for chondritic planetesimal formation.
Here, we propose two scenarios: changing the density evolution pathway of CAs to retain chondrules in CAs and escaping from the ejection barrier of planetesimal formation via the streaming instability before CAs lose their chondrules.

\subsubsection{Modification of the Density Evolution Pathway}

If chondrite parent bodies form via direct aggregation of CAs, then these aggregates need to avoid both the ejection barrier and the radial drift barrier.
Here, we can recognize the ``channel'' of the two barriers in Figures \ref{fig3} and \ref{fig4}.
If the pathway of density evolution passes the channel, then the chondrite parent bodies might form via direct aggregation.
Note that the density determined by static compression is the lower limit of the density evolution and that the bulk density of the CAs could increase if we nullify the assumption that CAs grow via collisions between similar-sized aggregates.
\citet{Dominik2009} showed that the hierarchical growth of large aggregates via the accretion of small aggregates leads to aggregates with a relatively dense structure.
In addition, aggregates formed via hierarchical growth can explain the small tensile strength inferred from the release of dust aggregates from the surface layers of cometary nuclei \citep{Skorov+2012}.
The in situ observation of dust particles from comet 67P/Churyumov-Gerasimenko also implies that hierarchical aggregation could be a dominant process for dust growth \citep{Bentley+2016}.
Although the formation processes of comets are not completely the same as those of chondrite parent bodies, we should consider the hierarchical growth of both MAs and CAs.

\subsubsection{Jumping over the Ejection Barrier}

It is expected that chondrite parent bodies can be formed via gravitationally bound clumps of CAs and that the formation of dust clumps is triggered by the streaming instability if the onset of the instability is faster than the onset of critical ejection.
As shown in Figures \ref{fig3} and \ref{fig4}, the critical ejection of chondrules from CAs starts when the Stokes number of the CAs ${\rm St}_{\rm CA}$ reaches of the order of $10^{-5}$--$10^{-3}$, depending on the disk parameters and the evaluation of the initial penetration velocity.
However, \citet{Carrera+2015} showed that whether the streaming instability occurs strongly depends on the Stokes number and that for a case where the dust-to-total mass ratio $Z$ is less than a few $10^{-2}$, the streaming instability can drive only when the Stokes number is within the range of $10^{-2} \lesssim {\rm St}_{\rm CA} \lesssim 1$.
We colored this region in the lower panels in Figures \ref{fig3} and \ref{fig4} (the yellow regions).
Therefore, forming chondrite parent bodies via the streaming instability might be improbable.

Note, however, that the critical Stokes number for driving the streaming instability is not yet fully understood, especially in the small Stokes number region \citep{Yang+2016}.
Moreover, there are several alternative models for making dust clumps, e.g., dust trapping at the local pressure maxima \citep[e.g.,][]{Drazkowska+2013} and/or vortices generated by hydrodynamical instabilities \citep[e.g.,][]{Meheut+2012,Gibbons+2015}.
Therefore, studies of these mechanisms taking fluffy growth and the retainment of chondrules into account are necessary.

\section{Summary}

Fluffy aggregation helps dust aggregates grow into planetesimals via direct aggregation.
Therefore, the chondrite parent bodies might have also formed via the fluffy aggregation of millimeter-sized chondrules and submicron-sized matrix grains (Figure \ref{fig1}).
However, chondrules can be ejected from the highly porous matrices when two chondrule-dust compound aggregates collide (Figure \ref{fig2}).

In this study, we introduced a density evolution model of chondrule-dust compound aggregates and calculated the evolutional track of porous aggregates.
In addition, we estimated the stopping length of the chondrules in the matrices.
For simplicity, we assumed that the chondrules are only decelerated by the drag force for penetration and the initial penetration velocity of the chondrules was parameterized due to the uncertainty in the collision behavior.

The main finding of this study is that fluffy aggregates likely lose their chondrules before they grow to meter-sized bodies, even if we evaluate the stopping length using lenient assessments (Figures \ref{fig3} and \ref{fig4}).
This ``ejection barrier'' might also prevent chondrule-dust compound aggregates from growing into planetesimals via the streaming instability.
Therefore, a change in the density evolution pathway is necessary to form chondrite parent bodies in our solar system.

\acknowledgments

The author thanks Hitoshi Miura, Taishi Nakamoto, Satoshi Okuzumi, and Hidekazu Tanaka for valuable comments.
The author is also thankful to the anonymous reviewer for constructive comments.
This work is supported by Grant-in-Aid for JSPS Research Fellow (17J06861).


%% for the paper.  Each keyword is check against the master list during
%% copy editing.  Individual instruments can be provided in parentheses,
%% after the keyword, but they are not verified.

%%\vspace{5mm}
%%\facilities{HST(STIS), Swift(XRT and UVOT), AAVSO, CTIO:1.3m, CTIO:1.5m,CXO}

%%\software{IRAF, cloudy, IDL}



\bibliographystyle{aasjournal}

%%\allauthors

%%\listofchanges

\onecolumngrid 
\listofchanges

\end{document}